\documentclass[sigconf]{acmart}
\renewcommand\footnotetextcopyrightpermission[1]{} 
\AtBeginDocument{%
  \providecommand\BibTeX{{%
    \normalfont B\kern-0.5em{\scshape i\kern-0.25em b}\kern-0.8em\TeX}}}

\acmDOI{}
\acmConference[Accepted at AICS 2022]{}{December,
  2022}{Ireland}

\usepackage{bm}
\usepackage{dirtytalk}

\newcommand\blfootnote[1]{%
  \begingroup
  \renewcommand\thefootnote{}\footnote{#1}%
  \addtocounter{footnote}{-1}%
  \endgroup
}
\begin{document}

\title[Multimodal Stock Embeddings]{A Multimodal Embedding-Based Approach to Industry Classification in Financial Markets*}

\author{Rian Dolphin}
\orcid{0000-0002-5607-9948}
\affiliation{%
  \institution{School of Computer Science, University College Dublin}
  \city{Dublin}
  \country{Ireland}
}
\email{rian.dolphin@ucdconnect.ie}

\author{Barry Smyth}
\orcid{0000-0003-0962-3362}
\affiliation{%
  \institution{School of Computer Science, University College Dublin}
  \city{Dublin}
  \country{Ireland}
}
\email{barry.smyth@ucd.ie}

\author{Ruihai Dong}
\orcid{0000-0002-2509-1370}
\affiliation{%
  \institution{School of Computer Science, University College Dublin}
  \city{Dublin}
  \country{Ireland}
}
\email{ruihai.dong@ucd.ie}

\renewcommand{\shortauthors}{R. Dolphin, et al.}

\begin{abstract}
Industry classification schemes provide a taxonomy for segmenting companies based on their business activities. They are relied upon in industry and academia as an integral component of many types of financial and economic analysis. However, even modern classification schemes have failed to embrace the era of big data and remain a largely subjective undertaking prone to inconsistency and misclassification. To address this, we propose a multimodal neural model for training company embeddings, which harnesses the dynamics of both historical pricing data and financial news to learn objective company representations that capture nuanced relationships. We explain our approach in detail and highlight the utility of the embeddings through several case studies and application to the downstream task of industry classification.
\end{abstract}




\maketitle\blfootnote{* Accepted at AICS 2022 under title \say{A Machine Learning Approach to Industry Classification in Financial Markets}. Preliminary version under this title was discussed at ICAIF '22 Workshop on NLP and Network Analysis in Financial Applications.}
\thispagestyle{empty}
\section{Introduction}
Financial markets are an important but challenging machine learning domain when it comes to analysis and prediction~\cite{bachelier1900theorie,fama1970efficient,fama1965behavior}. Their stochastic nature reflects a complex network of interactions involving a web of hidden factors and unpredictable events. Though the financial literature spans many sub-domains, the application of machine learning and deep learning techniques to financial markets has often been narrowly focused on the problem of returns forecasting for \emph{individual} assets~\cite{vachhani2019machine}. As a result, many other problems facing the financial sector have been underrepresented or ignored.

For example, the challenge of classifying companies based on a taxonomy of industry types is not well covered by contemporary machine learning research, even though it is an important task in several settings. In government, private sector, academia, and even the broader public, industry classification schemes are an integral part of using business and economic information~\cite{phillips2016industry}. Additionally, research has shown that 30\% of publications at the top-three finance journals utilize industry classification schemes~\cite{weiner2005impact}. 
The ability to segment companies into market sectors is important for many types of financial and economic analysis --- measuring economic activity, identifying peers and competitors, constructing ETF products, quantifying market share and bench-marking company performance --- none of which would be possible without industry classifications~\cite{phillips2016industry}. 

The rise in popularity of sector-based investing has led to the development of new market-oriented industry classification schemes. However, despite their increased usage for choosing investments, many industry classification schemes have still not embraced the era of big data and remain a largely subjective undertaking. As a result, they have been shown to struggle with scalability~\cite{phillips2016industry}, exhibit inconsistencies when determining the primary area of activity for a company~\cite{parker2012restoring}, and they offer no way to quantitatively measure or rank similarity between companies. Other studies confirm significant disagreement between classification schemes when trying to categorize the same companies~\cite{kahle1996impact, guenther1994differences}.

Although research applying modern computational techniques to industry classification schemes, and the learning of asset relationships more broadly, has lagged, there has been marked progress within the computer science community on relevant areas like representation learning and machine learning on relational data. The model architecture presented in this paper takes inspiration from a class of modern language models that have proven to be very successful in the natural language processing (NLP) domain \cite{mikolov2013efficient, pennington2014glove,firth1957synopsis}.

In this work, we propose a novel training methodology for the learning of distributed representations of public companies, based on distributional similarities in both historical returns data and financial news content. We show how these \emph{multimodal embeddings} can successfully capture the nuanced relationships that exist between companies and we demonstrate how they can be used to identify related companies. After discussing related work in the next section, we go on to explain the proposed approach before presenting several case-studies to highlight how the learned representations can be useful in financial applications. Before concluding, we present the results of an initial evaluation to demonstrate the effectiveness of these learned representations for the downstream task of industry classification.





\section{Related Work}



Having adopted machine learning in the 1980's, finance has long been a pioneering industry in the application of machine learning techniques~\cite{vachhani2019machine}.
Since then, interest has not waned and financial applications have remained a popular area of research in both academia and industry.
However, the literature applying modern computational techniques to financial markets overwhelmingly focuses on the forecasting of returns and volatility for \emph{individual} stocks~\cite{li2021modeling}.
Although these applications have seen success, there are many other tasks within financial markets which have not received the same level of attention. In this paper, we look to address one of these -- namely, the problem of industry classification.

Moving away from treating companies independently, and instead leveraging relationships, is key to tackling this task. Recent advances in areas such as representation learning and graph ML have encouraged research in this direction, with the most relevant literature to this work being papers proposing novel embedding frameworks for financial assets. For example,
\cite{wu2021equity2vec} suggest using matrix factorization to learn latent representations of stocks based on a co-occurrence matrix obtained from financial news articles.
The authors in \cite{satone2021fund2vec} use network theory and machine learning to generate fund and ETF embeddings based on overlapping asset allocation, \cite{sarmah2022learning} apply Node2Vec to the stock correlation matrix to learn embedded representations of stocks, and \cite{ang2021learning} obtain embeddings by combining company information with knowledge from Wikipedia and relationships from Wikidata. Most closely related to this work is \cite{dolphin2022stock}, where a framework is proposed for learning stock embeddings from the co-dynamics of historical returns data. This work directly builds on \cite{dolphin2022stock} by adding an additional data modality and focusing on the sector classification problem.

Other relevant applications of NLP in the financial domain include \cite{wan2021sentiment} where sentiment dynamics in related companies are assessed by building a network from financial news data.  Authors in \cite{kim2021artificial} tackle industry classification by using NLP to extract distinguishable features from business descriptions in financial reports.
Also, authors in \cite{ito2020embedding} extract company embeddings by using the output of the BERT~\cite{devlin2018bert} language model applied to annual reports, and then use these embeddings in the industry classification task.
Though textual data has been used to inform company embeddings in prior research, there is often a reliance on the aggregation of pre-trained word embeddings rather than a tailored company embedding framework, as proposed in this work.






\section{A Multimodal Embedding-Based Approach}\label{sec:methodology}

Applications of machine learning to financial forecasting tasks, particularly within the context of the stock market, have become a very popular area of research within the computer science community in recent years. However, this literature largely ignores the influence of groups of related companies, and instead treats companies as independent entities~\cite{li2021modeling}. This ignores an important source of relational information, which we believe can be encoded in dense vector representations of companies through training on multiple sources of data. 


\subsection{Language Modelling Origins}


This paper proposes a method for training embeddings of companies using a probabilistic neural framework. Inspired by the use of distributional semantics in natural language processing, we propose a model architecture that uses the idea of \emph{context companies} to train distributed representations of target companies. In linguistics, the distributional hypothesis captures the idea that \say{a word is characterized by the company it keeps}~\cite{firth1957synopsis}, i.e., words that occur in the same contexts tend to have similar meanings. In language modelling, the \emph{context} of any given word has quite a natural interpretation as the words immediately before and after it. 
However, defining \emph{context} in the case of financial assets is not as intuitive. But, before laying out the proposed approach to for financial assets, we first give some background on the Word2Vec~\cite{mikolov2013efficient} architecture to frame the discussion.

The distributional hypothesis underpins many modern language models like Word2Vec~\cite{mikolov2013efficient}. The goal of such language models is to construct a lower dimensional, dense representation of words that capture meaningful semantic and syntactic relationships~\cite{satone2021fund2vec}. The typical model architecture is a shallow two-layer neural network where the embeddings themselves are also the model parameters/weights. The embeddings are randomly initialized, as would be expected for weights in a neural model, and then trained by using \emph{context words} (the model input) to predict the \emph{center/target word} (the model output) and back propagating the loss\footnote{There are actually two Word2Vec architectures, we focus on CBOW and do not describe Skip-Gram here.}. This back propagation updates the embeddings in such a way that, after training, words which commonly appear in the same context will have similar representations in the latent space. Further information about this architecture and training process can be found in \cite{mikolov2013efficient, rong2014word2vec}.

To adapt this modelling framework to non-textual data, we must consider the idea of \emph{context} in the case of the two data sources we consider in this work: financial news and historical stock returns.
Section \ref{sec:num_intro} describes the selection of context companies from historical returns data, and Section \ref{sec:news_intro} outlines the same for financial news data. 
Then, Section \ref{sec:learning_process} describes the model architecture used to learn the embeddings in more detail, including examples from both the financial news and numerical returns data modalities.




\subsection{Selecting Context Companies from Historical Returns}\label{sec:num_intro}

\begin{figure*}
    \centering
    \includegraphics[width=0.75\linewidth]{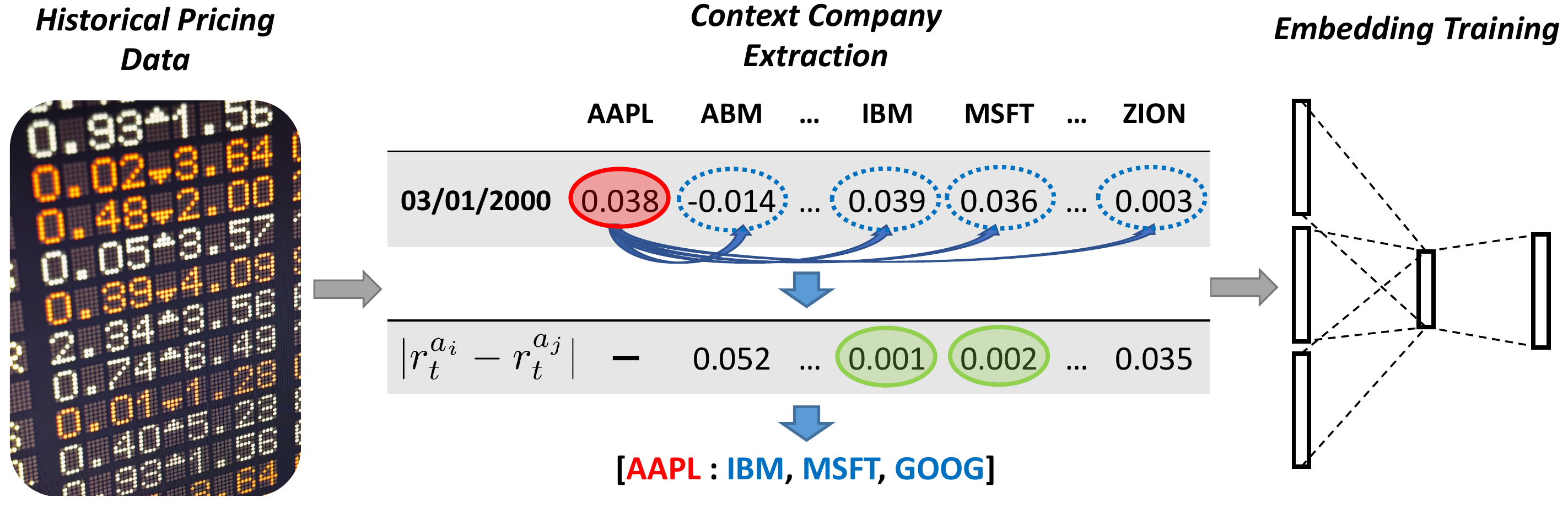}
    \caption{Embedding Pipeline for Historical Prices}
    \label{fig:numerical_architecture}
\end{figure*}

To select context companies from historical returns data, we treat companies with similar returns at the same points in time as related. This idea is supported by research showing that companies from the same business sectors tend to exhibit similar stock price fluctuations~\cite{gopikrishnan2000sector_price}. 




Consider a universe of public companies $U = \{a_1,...,a_{|U|}\}$ and for each company $a_i$ we have a vector $\mathbf{p}_{a_i}=\{p_0^{a_i},...,p_T^{a_i}\}$ containing its prices at discrete time intervals (daily or weekly for example). From the pricing data, we then compute a returns vector $\mathbf{r}_{a_i}=\{r_1^{a_i},...,r_T^{a_i}\}$ using Equation \ref{eqn:returns_formula}.
\begin{equation}
    \label{eqn:returns_formula}
    r_{t}^{a_i} = \frac{p_{t}^{a_i}-p_{t-1}^{a_i}}{p_{t-1}^{a_i}}
\end{equation}

We  generate \emph{target:context sets} from these returns vectors. For a context size $C$, the context companies for target asset $a_i$ at time $t$ are simply the $C$ companies which have the closest return at that point in time. The closest return is defined by the lowest absolute value difference in return for candidate company $a_j$, formulated as $|r_t^{a_i} - r_t^{a_j}|$. An example of this process is outlined in Figure \ref{fig:numerical_architecture} with Apple Inc. (AAPL) as the context company and $t$ as January 3\textsuperscript{rd} 2000. We compute the absolute value difference between the return of AAPL on that day with the return of each other company on the same day. Then, we choose the $C$ companies with the lowest difference values as the context companies, excluding AAPL itself. More generally, we generate a target:context set for every company at each point in time, which results in a total of $|U|\times T$ sets for training.

An example of a target:context set for $C=3$ might be \emph{[MSFT : IBM, AAPL, ORCL]}. This tells us that, at some point in time, the three companies with the closest returns to Microsoft were IBM, Apple Inc. and Oracle.

Market data is notoriously noisy~\cite{de1990noise}, and when looking at returns on the daily level, we see a bell shaped curve around that day's market average. This means that if the target stock has a return close to the market average on a given day, a lot of the corresponding context stocks are likely due to random chance. In an effort to isolate meaningful cases and reduce noise, a context set $\mathcal{S}(a_i,t)$, with target company $a_i$ at time $t$, was deleted from the training data if the target stock return, $r_t^{a_i}$, was within the interquartile range (IQR) of returns on that day. As a result, only sets where the target stock had a movement outside the IQR of market returns on a given day were included in training.

The remaining target:context sets are then passed into the embedding training architecture, which will be described in more detail in Section \ref{sec:learning_process}.

\subsection{Selecting Context Companies from News Articles}\label{sec:news_intro}

\begin{figure}
    \centering
    \includegraphics[width=1\linewidth]{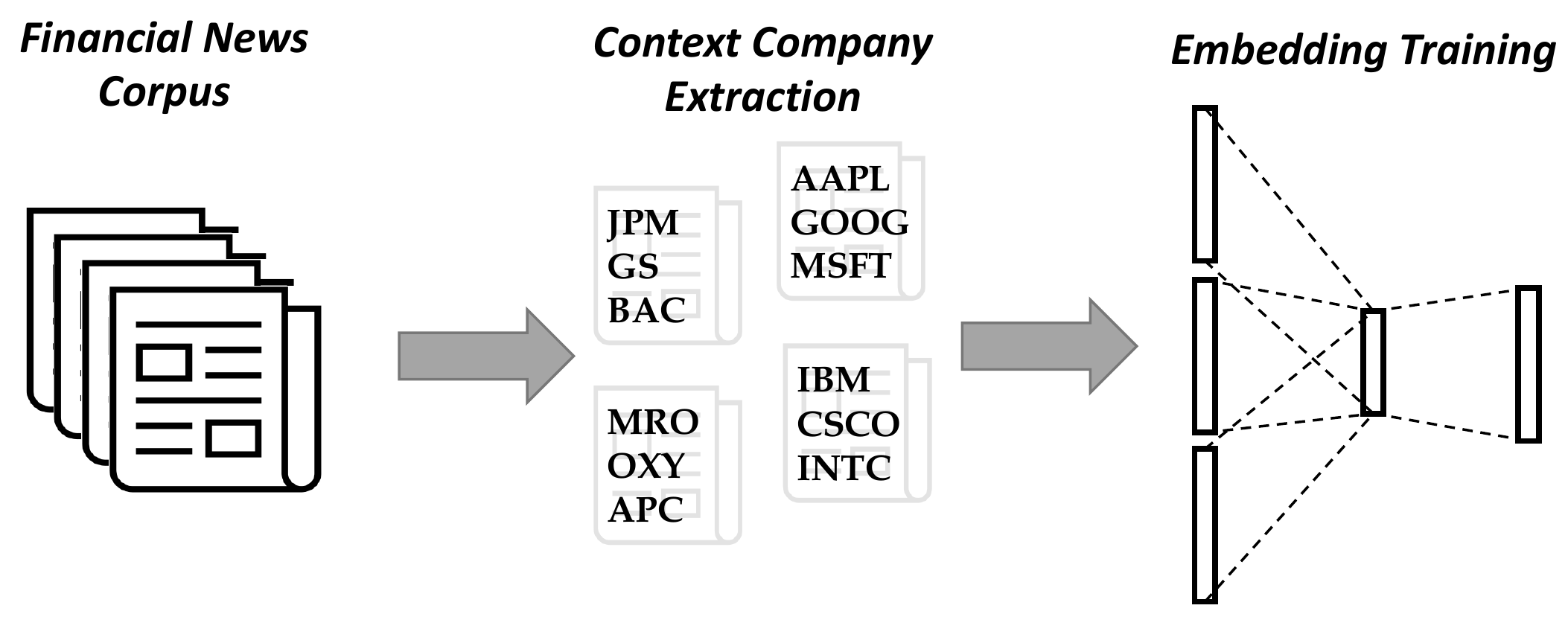}
    \caption{Embedding Pipeline for Financial News}
    \label{fig:news_architecture}
\end{figure}


Unsurprisingly, movements in financial markets and financial news have been found to be intrinsically linked. For example, a positive correlation exists between the number of occurrences of a company in the Financial Times on a given day and the transaction volume of that company’s stock both on the day before and the same day as the news is released~\cite{alanyali2013quantifying}. Additionally, with the evolution of NLP techniques, the application of language modelling to financial news data for stock market forecasting has become a popular area of research in recent years~\cite{yang2018explainable}. However, as previously mentioned, much of the research applying NLP techniques to financial news data has focused on forecasting and volatility prediction by using techniques such as sentiment analysis on a company by company basis~\cite{yang2018explainable}.


Focusing on individual companies/assets in isolation can mean that important relational information is missed. Here, we hypothesize that companies co-mentioned in the same news articles are likely to be related~\cite{wu2021equity2vec}, and that this can be leveraged to improve performance in tasks such as industry classification. As such, we want to learn embeddings in a way whereby companies which are commonly mentioned in the same news articles will end up having similar latent embeddings in terms of some suitable similarity metric, like cosine similarity for example. To do this, we use a corpus of over 100,000 financial news articles spanning 2006-2013~\cite{ding2014using}, and create target:context sets from every news article where more than one company is mentioned. 

As an example, if a given news article mentions $n$ companies, then each company will be put in a set as the target company with the remaining $n-1$ companies listed as context companies in that set. Therefore, an article with $n$ companies will result in $n$ target:context sets for training. These sets are then all passed into the model framework, where the embeddings are trained. This process is shown in Figure \ref{fig:news_architecture} and more detail will be given on the embedding training in Section \ref{sec:learning_process}.

\begin{figure}
    \centering
    \includegraphics[width=1\linewidth]{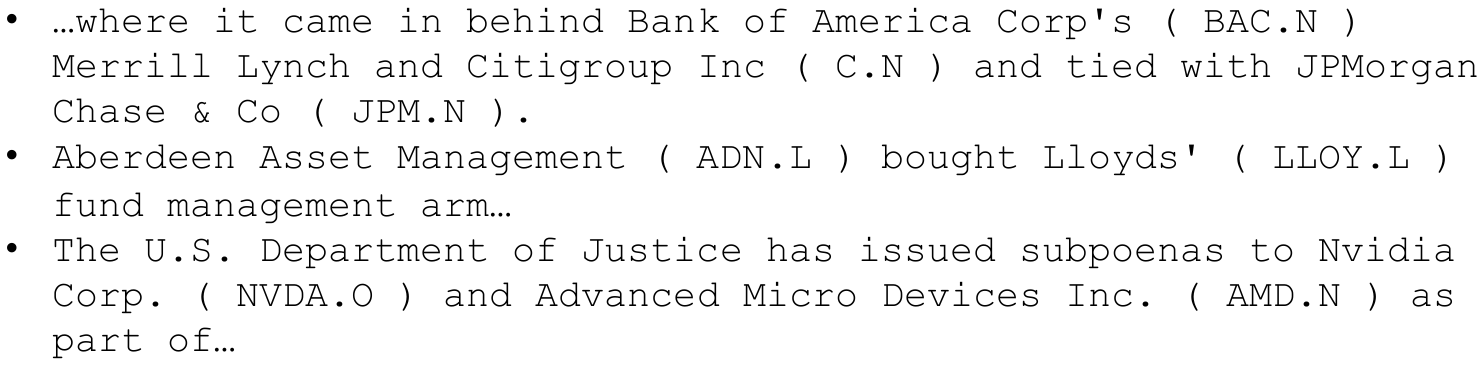}
    \caption{Example of News Snippets Showing Stock Tickers}
    \label{fig:news_examples}
\end{figure}

To carry out the process shown in Figure \ref{fig:news_architecture}, identifying when companies are mentioned in news articles is vital. This task is made easier because the news articles contain stock tickers where a publically listed company is mentioned. Figure \ref{fig:news_examples} shows snippets taken from the financial news corpus, and we can see that when a company is mentioned, it is also accompanied by its stock ticker enclosed in parentheses. For example, JPMorgan Chase \& Co is followed by (JPM.N). Tickers are consistently formatted throughout the dataset and thus can be extracted using regular expressions.


\subsection{The Training Process}\label{sec:learning_process}

\begin{figure}
    \centering
    \includegraphics[width=0.6\columnwidth]{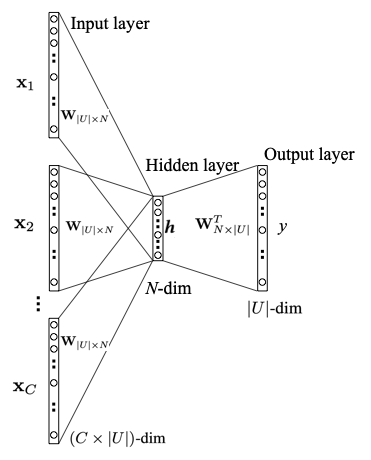}
    \caption{Model Architecture}
    \label{fig:CBOS}
\end{figure}

The aforementioned shallow two-layer neural network model architecture is illustrated in Figure \ref{fig:CBOS}. As previously mentioned, the model design is such that \emph{the company embeddings are the model parameters}. As such, each row in the weight matrix $\mathbf{W}$ is a company embedding. In this section, we explain each step of this framework in detail and why the resulting embeddings capture the relationships of interest. Throughout this section, it is worth keeping in mind that two of these models will be used -- one to learn the company embeddings for each data modality separately. The two independent embeddings are then concatenated to form the multimodal company embeddings.

The first step is to compute the hidden layer, which is simply an element-wise average of the context stock embeddings. To be more precise, the input to the model is a one-hot encoded version of the context set, and so, consists of $C$ one-hot vectors $\{\mathbf{x}_1,\mathbf{x}_2, ... , \mathbf{x}_C\}$, one for each context stock. These vectors are used to extract the embeddings corresponding to the $C$ context stocks. For example, computing $\mathbf{W}^T\cdot\mathbf{x}_1$ will extract a single row from $\mathbf{W}$ --- the embedding corresponding to the first context stock. The hidden layer, $\bm{h}$, is a simple element-wise average of the extracted embeddings, and is formulated in Equation \ref{eqn:hidden_layer}. We note that the use of a relatively simple average here is by design, since the mean function is agnostic to the number of inputs and so allows flexible context sizes during training.




\begin{equation}
    \label{eqn:hidden_layer}
    \bm{h} = \frac{1}{C} \mathbf{W}^T (\mathbf{x}_1+\mathbf{x}_2+ ... + \mathbf{x}_C)
\end{equation}

Thus, the hidden layer, $\bm{h}$, is an $N$-dimensional vector and can be thought of as an aggregate embedding representation of the context stocks, where $N$ is the embedding dimensionality. The next step is to estimate the probability of the target company \emph{given} $\bm{h}$ by applying Equation \ref{eqn:CBOW_out}.

\begin{equation}
    \label{eqn:CBOW_out}
    \mathbb{P}(\text{Target }|\text{ Context}) = softmax(\mathbf{W} \bm{h})
\end{equation}

Ensured by using the softmax activation, the output is a posterior probability distribution expressing the probability of each stock in the universe being the target stock given the context stocks observed. Since the dot product represents a measure of similarity between vectors, the model assigns higher probability to stocks whose embeddings are similar to hidden layer embedding $\bm{h}$. In this way, when we apply back-propagation, stocks which commonly co-occur in target:context sets will end up closer in the embedding space. As a result, assuming our hypotheses are correct, the embeddings will capture nuanced relationships that are present in the historical returns data and financial news co-occurrence. We note that the ground truth here, $y$ in Figure \ref{fig:CBOS}, is a one-hot vector indicating the true target stock.

\section{Evaluation}

In this section, we will outline a number of interesting example case studies followed by an evaluation on the task of industry classification. Understanding relationships between companies and segmenting them into industry sectors is vital for a wide range of applications in the financial domain~\cite{phillips2016industry} and we will describe some of these throughout this section.

\subsection{Datasets}
For this analysis, we use publically available daily historical pricing data and over 100,000 financial news articles~\cite{ding2014using}, both spanning 2006-2013. Included alongside the pricing data are two levels of industry labels for each stock, one high level like Finance, Technology etc., and the other a finer grained label like Major Bank or Semiconductors. The companies included in the analysis were selected based on the following inclusion criterion. Firstly, the company had to be publicly traded and have complete pricing data over the period in question. Secondly, we limited the dataset to companies mentioned in at least 50 news articles to ensure there was sufficient data for training. Additionally, the pricing data available contained only companies listed on the NYSE and NASDAQ exchanges, and so the dataset is also limited to these. After enforcing this inclusion criterion, we are left with 118 companies across seven industry sectors: Capital Goods, Consumer Non-Durables, Consumer Services, Energy, Finance, Health Care and Technology.

\subsection{Company Knowledge Graph}

\begin{figure*}[]
  \centering
  \includegraphics[width=1\linewidth]{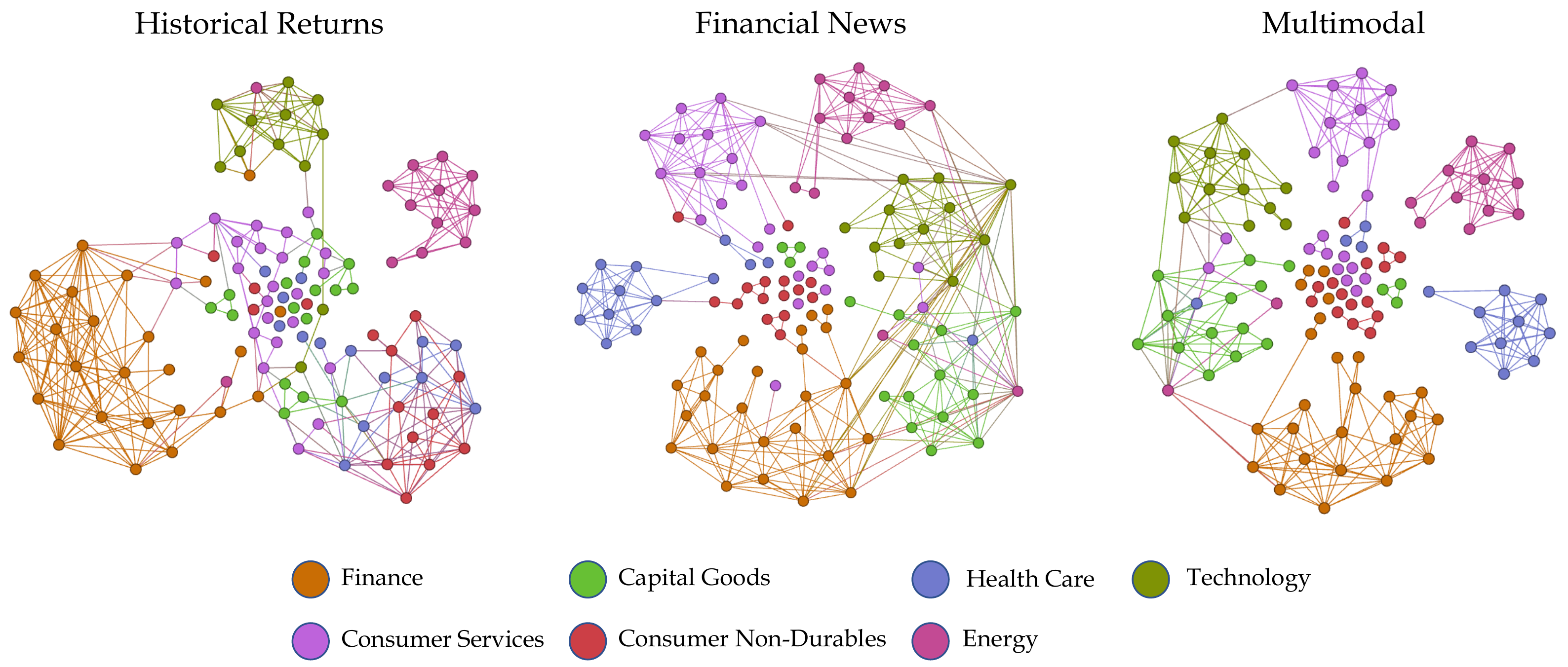}
  \caption{Visualization of Company Embeddings Colored by Industry Sector. Edges Indicate High Embedding Similarity.}
  \label{fig:triple_gephi}
\end{figure*}

Visualizing latent embeddings in two-dimensional space can often be a useful way of identifying relationships and clustering behavior. Figure \ref{fig:triple_gephi} shows three knowledge graphs, where each node represents a company and nodes are colored by their industry sector label, with seven industries in total. Each graph is derived from different embeddings: one from the historical return embeddings, one from the financial news-based embeddings and one constructed from a concatenation of both embeddings. It is worth noting that the dimensionality of both types of embeddings, a hyperparameter, was chosen to be 20. As a result, the concatenated multimodal embeddings are 40-dimensional.

Firstly, to convert embeddings into knowledge graphs, a similarity matrix was computed using the cosine similarity between company embeddings. Then, if two companies had a similarity above a certain threshold, they would receive an edge between their nodes. The similarity threshold was chosen as 0.6, which resulted in approximately 5\% of all possible edges being active. The plots are generated using a force-directed graph drawing algorithm in Gephi~\cite{ICWSM2009gephi}.

In each of the graphs in Figure \ref{fig:triple_gephi}, we observe clear clustering of companies into industry sectors. Within each graph, edges tend to be present between nodes of the same industry sector, though there are some exceptions which will be discussed further in Section \ref{sec:mismatch}. This indicates that the proposed training framework successfully learns embeddings which pick up on relationships between companies in the case of both data modalities. In each case, using returns data or news co-occurrence, this is a very positive result because it suggests that it is possible to reconstruct important sectoral information from the embeddings, and indeed is likely to do so in a way that is more nuanced and objective than might be possible using simple sectoral labels.

Contrasting the three knowledge graphs, the graph constructed from the combined embeddings seems to best cluster the companies into industry sectors. This indicates a benefit in combining the embeddings from both data modalities, which will be explored in further evaluations.


\subsection{Identifying Related Companies}\label{sec:KNN}


This first case study looks to use the learned embeddings to identify related companies through a nearest neighbors analysis, a natural first point of reference to sanity check the latent space representations. We would hope that companies with very high similarity in the latent space should be related somehow. In order to find the $k$-nearest neighbors (kNN) for a given query company, we first must define what exactly we mean by nearest. We implement kNN using cosine similarity as the similarity metric; note that a similar pattern of related companies results if we use euclidean distance or dot product similarity instead. 

\begin{table*}[]
    \centering
    \caption{Examples of Top-3 Nearest Neighbors for Given Query Companies}
    \label{tab:KNN}
    \begin{tabular}{cccc}
        \textbf{\begin{tabular}[c]{@{}c@{}c@{}}Query company\\ Industry 1 \\ Industry 2\end{tabular}} & \textbf{Neighbor Company - Industry 1 - Industry 2} & \textbf{Similarity}  \\ \midrule
        \begin{tabular}[c]{@{}c@{}}JPMorgan Chase\\ Finance \\Major Bank\end{tabular}          & \begin{tabular}[c]{@{}c@{}}Citibank - Finance - Major Bank \\ Bank of America Corp - Finance - Major Bank  \\ Wells Fargo \& Company - Finance - Major Bank\end{tabular}  & \begin{tabular}[c]{@{}c@{}c@{}}0.96\\0.95\\0.92\end{tabular} \\\hline
        
        \begin{tabular}[c]{@{}c@{}}Intel Corporation\\ Technology \\ Semiconductors\end{tabular}          & \begin{tabular}[c]{@{}c@{}}Apple Inc. - Technology - Computer Manufacturing \\ Texas Instruments - Technology - Semiconductors  \\ Hewlett-Packard - Technology - Computer Manufacturing \end{tabular} & \begin{tabular}[c]{@{}c@{}c@{}} 0.84 \\ 0.83 \\ 0.83\end{tabular} \\ \hline
        
        \begin{tabular}[c]{@{}c@{}}Walmart\\ Consumer Services \\ Department Stores\end{tabular}          & \begin{tabular}[c]{@{}c@{}}Target - Consumer Services - Department Stores \\ Costco -  Energy - Department Stores \\ Best Buy - Consumer Services - Consumer Electronics \end{tabular} & \begin{tabular}[c]{@{}c@{}c@{}}0.90\\0.85\\0.78\end{tabular}  \\ \hline
    \end{tabular}
\end{table*}

Table \ref{tab:KNN} shows the top-3 ($k=3$) nearest neighbors for JPMorgan Chase, Intel Corporation, and Walmart, three well-known companies in very different sectors. In each case, the nearest neighbors pass the \say{sanity test} in that they belong to similar industry sectors and in many cases also agree on the finer-grained classification labelled as \say{Industry 2} in Table \ref{tab:KNN}. For example, the three nearest neighbors of JPMorgan Chase, a major bank, are also all major banks. Remember, that no sectoral or industry information has been used in determining these nearest neighbors, and only daily returns and co-occurence in news articles have been used to generate the distributed representations used for similarity assessment.

There is considerable scope for the use of nearest neighbor companies by investors. Firstly, we can develop a company recommendation system which, when given a target company -- a novel company for the investor or one already in their portfolio -- can generate a ranked list of similar companies based on their historical returns data and appearance in financial news. A system like this addresses a major pitfall of classic industry classification schemes, where no rank ordering is possible. This could have a variety of use cases, for example, investors and fund managers could consult this ranked list when conducting comparable company analysis or looking for alternative investment opportunities; it could be of use to sales representatives looking to recommend complementary investment opportunities to clients; investors could devise a tax loss harvesting strategy~\cite{satone2021fund2vec}; and asset managers could be assisted in the construction of market sector ETFs.

\subsection{Analyzing High Similarity Mismatches}\label{sec:mismatch}

\begin{table*}[]
    \centering
    \caption{Examples of High Similarity Mismatches --- Companies With Very High Similarity That Have Different Sector Labels}
    \label{tab:mismatch}
    \begin{tabular}{ccc}
        \textbf{\begin{tabular}[c]{@{}c@{}}Company A \\ Industry 1 - Industry 2\end{tabular}}                         & \textbf{\begin{tabular}[c]{@{}c@{}}Company B\\ Industry 1 - Industry 2\end{tabular}}                     & \textbf{Similarity} \\ \midrule
        \begin{tabular}[c]{@{}c@{}}General Electric\\ Energy - Consumer Electronics\end{tabular}          & \begin{tabular}[c]{@{}c@{}}Boeing\\ Capital Goods - Aerospace\end{tabular}                   & 0.87                \\ \hline
        \begin{tabular}[c]{@{}c@{}}Johnson \& Johnson\\ Consumer Non-Durables - Cosmetics\end{tabular} & \begin{tabular}[c]{@{}c@{}}Colgate-Palmolive\\ Healthcare - Major Pharma\end{tabular}        & 0.81                \\ \hline
        \begin{tabular}[c]{@{}c@{}}3M Company\\ Healthcare - Medical Instruments\end{tabular}              & \begin{tabular}[c]{@{}c@{}}Honeywell International\\ Capital Goods - Auto Parts\end{tabular} & 0.88                \\ \hline
    \end{tabular}
\end{table*}

Though the vast majority of edges in Figure \ref{fig:triple_gephi} occur between nodes from the same industry, it is not true in all cases. In other words, there are some instances where two companies have a high embedding similarity, but their industry sector labels don't match. Does this highlight a flaw in the embeddings, where companies achieve very high embedding similarity when they should not? To answer this, we provide some examples of these \emph{high similarity mismatches}. We consider pairs of companies that have a high cosine similarity between embeddings and are members of different industry sectors. A number of examples are shown in Table \ref{tab:mismatch}.

The first example is that of General Electric, classified in the Energy (Consumer Electronics) sector, and Boeing, classified in the Capital Goods (Aerospace) sector. These two companies receive a high multimodal embedding similarity of 0.8 despite being classified in different industry sectors. However, upon closer inspection, we note that one of General Electric's main business areas is the manufacturing of aircraft engines, with Boeing being one of their largest customers. As a result, they are commonly mentioned in news articles with Boeing and other aerospace companies, which results in the high embedding similarity. It is interesting to note here that the returns embedding similarity between these companies is not as high (0.30) as their news embedding similarity (0.87). Perhaps this is because General Electric has many other business areas which influence its stock price fluctuations aside from its aerospace operations, and this prevents high returns-based similarity. However, the supply chain relationship between the two may result in news co-mentions, which increases the news embedding similarity. This interesting example indicates the potential for each data modality to pick up on different relationship types.

Another example is that of Johnson \& Johnson and Colgate-Palmolive, classified as Consumer Non-Durables (Cosmetics) and Healthcare (Major Pharma) respectively. These two companies again have a high embedding similarity, but have been classified into different industry sectors. The relationship could be explained by the large presence both companies have in the consumer healthcare market, resulting in exposure to similar idiosyncratic risk factors and the resulting headwinds. In fact, one could certainly make the case for both companies to be in either sector, which highlights the subjective methodology of current industry classification schemes and the useful ability of higher dimensional embeddings to pick up on more nuanced relationships.

The final example in Table \ref{tab:mismatch} is Honeywell and 3M. Again, they have relatively high similarity in their multimodal company embeddings, despite being classified in different industry sectors. The two companies are both multinational conglomerates operating in similar business sectors. In addition to this, both are constituents of three of the most popular indexes (Dow Jones Industrial Average, S\&P 500 and S\&P 100), and index inclusion has been shown to lead to more frequent news mentions and greater co-movement in returns~\cite{barberis2005comovement}.

Through these examples, we can see that current industry classification schemes will often segment quite similar companies into different industry sectors. The line is not always clear and, despite the increased usage of industry classification schemes for choosing investments, many have not embraced the era of big data and fail to utilize the countless data points being generated each day. Instead, the company allocation procedure remains a largely subjective task~\cite{phillips2016industry}. Using multimodal embeddings, on the other hand, removes subjectivity and allows the user to leverage the plethora of financial data continuously being generated. We propose that the embeddings could be a useful and objective tool to complement the existing methods by which companies are grouped. For example, a company which has many high similarity mismatches might be getting classified inconsistently to its peers, and would be worth further consideration.


\subsection{Using Multimodal Embeddings for Industry Classification}\label{sec:classification}

\begin{figure*}
    \centering
    \includegraphics[width=\linewidth]{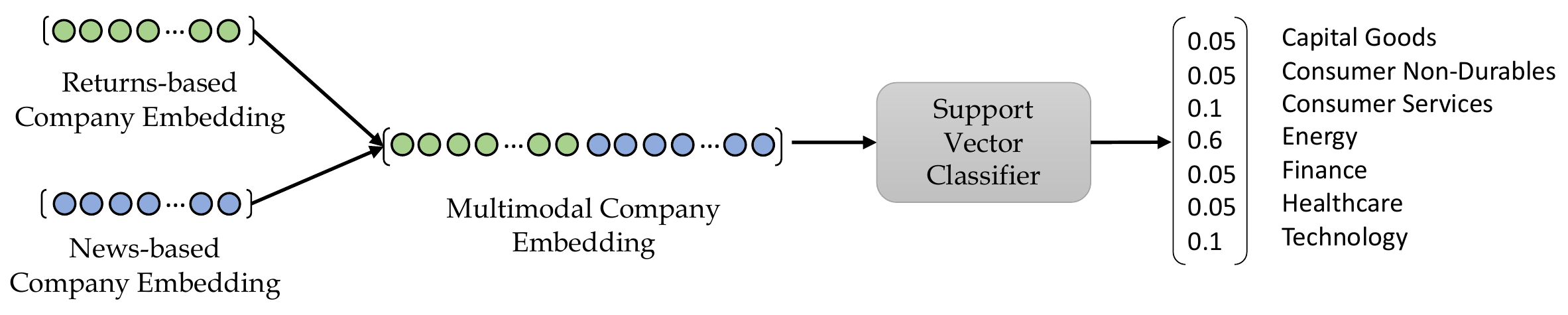}
    \caption{Classifier Framework}
    \label{fig:classifier}
\end{figure*}

Using the embeddings generated by the proposed training framework, we can use a classification model to segment companies into business sectors in an objective manner. To do this, we train a classification model with embeddings as the input and industry sector label as the output. As a classification model, we use a support vector classifier~\cite{platt1999SVM}, and to train the model, we used $k$-fold cross validation with $k=4$. Within the data there is a class imbalance between industry sectors which can introduce algorithmic bias, and so, we applied Synthetic Minority Oversampling Technique (SMOTE) to the training data to account for this.

There are a number of considerations here which will undoubtedly limit the accuracy of the classification model. Firstly, the embeddings themselves are derived solely from historical returns data and financial news, which are both influenced by a complex network of unpredictable factors. Secondly, as we saw in Section \ref{sec:mismatch} when looking at high similarity mismatches, there are a number of companies with subjective ground truth labels that could be placed in a number of the industry categories. As a result, industry classification in the financial domain is a challenging problem. 

Despite these hurdles, we see accuracy of 90\% when using the multimodal embeddings to classifying companies into industry sector. Also reported is the accuracy obtained from the returns-based embeddings and the news-based embeddings by themselves. We see an accuracy of 85\% when using the news-only embeddings, which is also above the baseline accuracy of 72\% from the returns-only method.

Table \ref{tab:hold_out_results} shows the results from a hold-out testing approach as opposed to the $k$-fold cross validation, and gives a more detailed breakdown of performance by industry sector. The results show particularly strong performance for many of the more highly populated industry sectors. Increasing the number of companies present in each industry segment would likely increase the generalizability of the model. In an effort to do this, we experimented with various thresholds for the number of news article mentions needed for company inclusion. By lowering this threshold (i.e., including companies even if they were mentioned in less than 50 news articles), the model accuracy decreases because of lower quality embeddings for companies with less data and an increase in the number of industry sectors. Sensitivity analysis led to the choice of 50 as the threshold.

\begin{table}[]
    \centering
    \caption{Results From Industry Classification Task Using $k$-Fold Cross Validation With $k=4$}\label{tab:classification_results}
    \begin{tabular}{ccccc}
    \toprule
        \textbf{Model}                               & \textbf{Precision} & \textbf{Recall} & \textbf{F1} & \textbf{Accuracy} \\
        \midrule
        Returns Embedding                                    & 0.74               & 0.72            & 0.72        & 72\%              \\
        News Embedding                              & 0.89               & 0.85            & 0.84        & 85\%              \\
        Multimodal Embedding                           & \textbf{0.91}               & \textbf{0.90}            & \textbf{0.90}        & \textbf{90\% }             \\
        \bottomrule
    \end{tabular}
\end{table}

\begin{table}[]
    \centering
    \caption{Breakdown of Industry Classification for Hold-Out Example}
    \label{tab:hold_out_results}
    \begin{tabular}{lccc}
        \toprule
        \textbf{Industry Class} &  \textbf{Precision} &  \textbf{Recall} &  \textbf{F1-Score} \\
        \midrule
        Capital Goods         &       0.80 &    1.00 &      0.89  \\
        Consumer Non-Durables &       1.00 &    0.60 &      0.75  \\
        Consumer Services     &       0.67 &    1.00 &      0.80  \\
        Energy                &       1.00 &    0.60 &      0.75  \\
        Finance               &       1.00 &    1.00 &      1.00  \\
        Health Care           &       1.00 &    1.00 &      1.00  \\
        Technology            &       1.00 &    1.00 &      1.00  \\
        \midrule
        Weighted Avg          &       0.92 &    0.89 &      0.88  \\
        \midrule
        Overall Accuracy              &        &     &      0.89 \\
        
        \bottomrule
    \end{tabular}
    \label{tab:full_classification_report}
\end{table}

A model allowing objective segmentation of companies into industry sectors as proposed here has a number of use cases. In particular, this model could provide useful assistance to companies offering ETF products, demand for which has grown rapidly in recent years. In fact, ETFs have become so popular that, by the end of 2016, their market share exceeded 10\% of the total market capitalization traded on US exchanges, representing more than 30\% of overall trading volume~\cite{ben2016exchange}.
Sometimes the portfolios are constructed to track certain indices, such as the S\&P500. However, the inclusion criterion for which stocks make it into a particular index/ETF can be highly subjective, particularly for ETFs designed to track certain industry sectors. For example, a company like Amazon is classified as \say{consumer discretionary} by the Global Industry Classification Standard (GICS). Should it be included in a consumer discretionary ETF with the likes of Ford Motor and McDonald's? Or would it be better suited to a technology or consumer staples ETF? Indeed, a case can be made for Amazon to be considered as all of the above.

A modelling tool like the one we have described outputs a probability that a given company is in each industry sector. This score could give analysts further indication of whether a company is suitable for inclusion in a particular index/ETF and augment the current subjective decision-making process.





\section{Conclusion}

This work has focused on leveraging multiple sources of data to tackle the industry classification problem using machine learning. We proposed an approach for learning dense vector representations of companies which capture nuanced and interesting relationships between companies. The potential utility of these embeddings to financial analysts was discussed in relation to a number of tasks, and the evaluation results speak to the potential benefits of our approach and provide a useful starting point for further exploration and development. From examples in Section \ref{sec:KNN}, we saw that the embeddings trained on each modality picked up on interesting relationships of different types. The benefits of the multimodal approach were also highlighted in the results of the industry classification model in Section \ref{sec:classification}, where combining modalities increased overall accuracy and F1 score.

In future work, we plan to adapt the proposed framework to generate multimodal embeddings optimized to capture dissimilarity, in addition to similarity, which is an important tool for effective portfolio optimization. We believe that these embeddings have the potential to be useful in the asset management space by informing successful diversification and risk management strategies. In addition, we plan to utilize other sources of data and introduce sentiment aware context company selection to assess the impact on performance. 


\textbf{Acknowledgements.} This publication has emanated from research conducted with the financial support of Science Foundation Ireland under Grant number 18/CRT/6183.

\bibliographystyle{ACM-Reference-Format}
\bibliography{bibliography_old, bibliography_new}


\begin{thebibliography}{32}


\ifx \showCODEN    \undefined \def \showCODEN     #1{\unskip}     \fi
\ifx \showDOI      \undefined \def \showDOI       #1{#1}\fi
\ifx \showISBNx    \undefined \def \showISBNx     #1{\unskip}     \fi
\ifx \showISBNxiii \undefined \def \showISBNxiii  #1{\unskip}     \fi
\ifx \showISSN     \undefined \def \showISSN      #1{\unskip}     \fi
\ifx \showLCCN     \undefined \def \showLCCN      #1{\unskip}     \fi
\ifx \shownote     \undefined \def \shownote      #1{#1}          \fi
\ifx \showarticletitle \undefined \def \showarticletitle #1{#1}   \fi
\ifx \showURL      \undefined \def \showURL       {\relax}        \fi
\providecommand\bibfield[2]{#2}
\providecommand\bibinfo[2]{#2}
\providecommand\natexlab[1]{#1}
\providecommand\showeprint[2][]{arXiv:#2}

\bibitem[Alanyali et~al\mbox{.}(2013)]%
        {alanyali2013quantifying}
\bibfield{author}{\bibinfo{person}{Merve Alanyali},
  \bibinfo{person}{Helen~Susannah Moat}, {and} \bibinfo{person}{Tobias Preis}.}
  \bibinfo{year}{2013}\natexlab{}.
\newblock \showarticletitle{Quantifying the relationship between financial news
  and the stock market}.
\newblock \bibinfo{journal}{\emph{Scientific reports}} \bibinfo{volume}{3},
  \bibinfo{number}{1} (\bibinfo{year}{2013}), \bibinfo{pages}{1--6}.
\newblock


\bibitem[Ang and Lim(2021)]%
        {ang2021learning}
\bibfield{author}{\bibinfo{person}{Gary Ang} {and} \bibinfo{person}{Ee-Peng
  Lim}.} \bibinfo{year}{2021}\natexlab{}.
\newblock \showarticletitle{Learning knowledge-enriched company embeddings for
  investment management}. In \bibinfo{booktitle}{\emph{Proceedings of the
  Second ACM International Conference on AI in Finance}}.
  \bibinfo{pages}{1--9}.
\newblock


\bibitem[Bachelier(1900)]%
        {bachelier1900theorie}
\bibfield{author}{\bibinfo{person}{Louis Bachelier}.}
  \bibinfo{year}{1900}\natexlab{}.
\newblock \showarticletitle{Th{\'e}orie de la sp{\'e}culation}. In
  \bibinfo{booktitle}{\emph{Annales scientifiques de l'{\'E}cole normale
  sup{\'e}rieure}}, Vol.~\bibinfo{volume}{17}. \bibinfo{pages}{21--86}.
\newblock


\bibitem[Barberis et~al\mbox{.}(2005)]%
        {barberis2005comovement}
\bibfield{author}{\bibinfo{person}{Nicholas Barberis}, \bibinfo{person}{Andrei
  Shleifer}, {and} \bibinfo{person}{Jeffrey Wurgler}.}
  \bibinfo{year}{2005}\natexlab{}.
\newblock \showarticletitle{Comovement}.
\newblock \bibinfo{journal}{\emph{Journal of financial economics}}
  \bibinfo{volume}{75}, \bibinfo{number}{2} (\bibinfo{year}{2005}),
  \bibinfo{pages}{283--317}.
\newblock


\bibitem[Bastian et~al\mbox{.}(2009)]%
        {ICWSM2009gephi}
\bibfield{author}{\bibinfo{person}{Mathieu Bastian}, \bibinfo{person}{Sebastien
  Heymann}, {and} \bibinfo{person}{Mathieu Jacomy}.}
  \bibinfo{year}{2009}\natexlab{}.
\newblock \bibinfo{title}{Gephi: An Open Source Software for Exploring and
  Manipulating Networks}.
\newblock
\newblock
\urldef\tempurl%
\url{http://www.aaai.org/ocs/index.php/ICWSM/09/paper/view/154}
\showURL{%
\tempurl}


\bibitem[Ben-David et~al\mbox{.}(2016)]%
        {ben2016exchange}
\bibfield{author}{\bibinfo{person}{Itzhak Ben-David},
  \bibinfo{person}{Francesco Franzoni}, {and} \bibinfo{person}{Rabih
  Moussawi}.} \bibinfo{year}{2016}\natexlab{}.
\newblock \bibinfo{booktitle}{\emph{Exchange traded funds (ETFs)}}.
\newblock \bibinfo{type}{{T}echnical {R}eport}. \bibinfo{institution}{National
  Bureau of Economic Research}.
\newblock


\bibitem[De~Long et~al\mbox{.}(1990)]%
        {de1990noise}
\bibfield{author}{\bibinfo{person}{J~Bradford De~Long}, \bibinfo{person}{Andrei
  Shleifer}, \bibinfo{person}{Lawrence~H Summers}, {and}
  \bibinfo{person}{Robert~J Waldmann}.} \bibinfo{year}{1990}\natexlab{}.
\newblock \showarticletitle{Noise trader risk in financial markets}.
\newblock \bibinfo{journal}{\emph{Journal of political Economy}}
  \bibinfo{volume}{98}, \bibinfo{number}{4} (\bibinfo{year}{1990}),
  \bibinfo{pages}{703--738}.
\newblock


\bibitem[Devlin et~al\mbox{.}(2018)]%
        {devlin2018bert}
\bibfield{author}{\bibinfo{person}{Jacob Devlin}, \bibinfo{person}{Ming-Wei
  Chang}, \bibinfo{person}{Kenton Lee}, {and} \bibinfo{person}{Kristina
  Toutanova}.} \bibinfo{year}{2018}\natexlab{}.
\newblock \showarticletitle{Bert: Pre-training of deep bidirectional
  transformers for language understanding}.
\newblock \bibinfo{journal}{\emph{arXiv preprint arXiv:1810.04805}}
  (\bibinfo{year}{2018}).
\newblock


\bibitem[Ding et~al\mbox{.}(2014)]%
        {ding2014using}
\bibfield{author}{\bibinfo{person}{Xiao Ding}, \bibinfo{person}{Yue Zhang},
  \bibinfo{person}{Ting Liu}, {and} \bibinfo{person}{Junwen Duan}.}
  \bibinfo{year}{2014}\natexlab{}.
\newblock \showarticletitle{Using structured events to predict stock price
  movement: An empirical investigation}. In
  \bibinfo{booktitle}{\emph{Proceedings of the 2014 conference on empirical
  methods in natural language processing (EMNLP)}}.
  \bibinfo{pages}{1415--1425}.
\newblock


\bibitem[Dolphin et~al\mbox{.}(2022)]%
        {dolphin2022stock}
\bibfield{author}{\bibinfo{person}{Rian Dolphin}, \bibinfo{person}{Barry
  Smyth}, {and} \bibinfo{person}{Ruihai Dong}.}
  \bibinfo{year}{2022}\natexlab{}.
\newblock \showarticletitle{Stock Embeddings: Learning Distributed
  Representations for Financial Assets}.
\newblock \bibinfo{journal}{\emph{arXiv preprint arXiv:2202.08968}}
  (\bibinfo{year}{2022}).
\newblock


\bibitem[Fama(1965)]%
        {fama1965behavior}
\bibfield{author}{\bibinfo{person}{Eugene~F Fama}.}
  \bibinfo{year}{1965}\natexlab{}.
\newblock \showarticletitle{The behavior of stock-market prices}.
\newblock \bibinfo{journal}{\emph{The journal of Business}}
  \bibinfo{volume}{38}, \bibinfo{number}{1} (\bibinfo{year}{1965}),
  \bibinfo{pages}{34--105}.
\newblock


\bibitem[Firth(1957)]%
        {firth1957synopsis}
\bibfield{author}{\bibinfo{person}{John~R Firth}.}
  \bibinfo{year}{1957}\natexlab{}.
\newblock \showarticletitle{A synopsis of linguistic theory, 1930-1955}.
\newblock \bibinfo{journal}{\emph{Studies in linguistic analysis}}
  (\bibinfo{year}{1957}).
\newblock


\bibitem[Gopikrishnan et~al\mbox{.}(2000)]%
        {gopikrishnan2000sector_price}
\bibfield{author}{\bibinfo{person}{Parameswaran Gopikrishnan},
  \bibinfo{person}{Bernd Rosenow}, \bibinfo{person}{Vasiliki Plerou}, {and}
  \bibinfo{person}{H~Eugene Stanley}.} \bibinfo{year}{2000}\natexlab{}.
\newblock \showarticletitle{Identifying business sectors from stock price
  fluctuations}.
\newblock \bibinfo{journal}{\emph{arXiv preprint cond-mat/0011145}}
  (\bibinfo{year}{2000}).
\newblock


\bibitem[Guenther and Rosman(1994)]%
        {guenther1994differences}
\bibfield{author}{\bibinfo{person}{David~A Guenther} {and}
  \bibinfo{person}{Andrew~J Rosman}.} \bibinfo{year}{1994}\natexlab{}.
\newblock \showarticletitle{Differences between COMPUSTAT and CRSP SIC codes
  and related effects on research}.
\newblock \bibinfo{journal}{\emph{Journal of Accounting and Economics}}
  \bibinfo{volume}{18}, \bibinfo{number}{1} (\bibinfo{year}{1994}),
  \bibinfo{pages}{115--128}.
\newblock


\bibitem[Ito et~al\mbox{.}(2020)]%
        {ito2020embedding}
\bibfield{author}{\bibinfo{person}{Tomoki Ito}, \bibinfo{person}{Jose
  Camacho~Collados}, \bibinfo{person}{Hiroki Sakaji}, {and}
  \bibinfo{person}{Steven Schockaert}.} \bibinfo{year}{2020}\natexlab{}.
\newblock \showarticletitle{Learning company embeddings from annual reports for
  fine-grained industry characterization}.
\newblock  (\bibinfo{year}{2020}).
\newblock


\bibitem[Kahle and Walkling(1996)]%
        {kahle1996impact}
\bibfield{author}{\bibinfo{person}{Kathleen~M Kahle} {and}
  \bibinfo{person}{Ralph~A Walkling}.} \bibinfo{year}{1996}\natexlab{}.
\newblock \showarticletitle{The impact of industry classifications on financial
  research}.
\newblock \bibinfo{journal}{\emph{Journal of financial and quantitative
  analysis}} \bibinfo{volume}{31}, \bibinfo{number}{3} (\bibinfo{year}{1996}),
  \bibinfo{pages}{309--335}.
\newblock


\bibitem[Kim et~al\mbox{.}(2021)]%
        {kim2021artificial}
\bibfield{author}{\bibinfo{person}{Daejin Kim}, \bibinfo{person}{Hyoung-Goo
  Kang}, \bibinfo{person}{Kyounghun Bae}, {and} \bibinfo{person}{Seongmin
  Jeon}.} \bibinfo{year}{2021}\natexlab{}.
\newblock \showarticletitle{An artificial intelligence-enabled industry
  classification and its interpretation}.
\newblock \bibinfo{journal}{\emph{Internet Research}} \bibinfo{volume}{32},
  \bibinfo{number}{2} (\bibinfo{year}{2021}), \bibinfo{pages}{406--424}.
\newblock


\bibitem[Li et~al\mbox{.}(2021)]%
        {li2021modeling}
\bibfield{author}{\bibinfo{person}{Wei Li}, \bibinfo{person}{Ruihan Bao},
  \bibinfo{person}{Keiko Harimoto}, \bibinfo{person}{Deli Chen},
  \bibinfo{person}{Jingjing Xu}, {and} \bibinfo{person}{Qi Su}.}
  \bibinfo{year}{2021}\natexlab{}.
\newblock \showarticletitle{Modeling the stock relation with graph network for
  overnight stock movement prediction}. In
  \bibinfo{booktitle}{\emph{Proceedings of the Twenty-Ninth International
  Conference on International Joint Conferences on Artificial Intelligence}}.
  \bibinfo{pages}{4541--4547}.
\newblock


\bibitem[Malkiel and Fama(1970)]%
        {fama1970efficient}
\bibfield{author}{\bibinfo{person}{Burton~G Malkiel} {and}
  \bibinfo{person}{Eugene~F Fama}.} \bibinfo{year}{1970}\natexlab{}.
\newblock \showarticletitle{Efficient capital markets: A review of theory and
  empirical work}.
\newblock \bibinfo{journal}{\emph{The journal of Finance}}
  \bibinfo{volume}{25}, \bibinfo{number}{2} (\bibinfo{year}{1970}),
  \bibinfo{pages}{383--417}.
\newblock


\bibitem[Mikolov et~al\mbox{.}(2013)]%
        {mikolov2013efficient}
\bibfield{author}{\bibinfo{person}{Tomas Mikolov}, \bibinfo{person}{Kai Chen},
  \bibinfo{person}{Greg Corrado}, {and} \bibinfo{person}{Jeffrey Dean}.}
  \bibinfo{year}{2013}\natexlab{}.
\newblock \showarticletitle{Efficient estimation of word representations in
  vector space}.
\newblock \bibinfo{journal}{\emph{arXiv preprint arXiv:1301.3781}}
  (\bibinfo{year}{2013}).
\newblock


\bibitem[Parker(2012)]%
        {parker2012restoring}
\bibfield{author}{\bibinfo{person}{RP Parker}.}
  \bibinfo{year}{2012}\natexlab{}.
\newblock \showarticletitle{Restoring the enterprise statistics program (ESP)
  for the 2012 economic census}.
\newblock \bibinfo{journal}{\emph{Reports of the Census Bureau}}
  (\bibinfo{year}{2012}).
\newblock


\bibitem[Pennington et~al\mbox{.}(2014)]%
        {pennington2014glove}
\bibfield{author}{\bibinfo{person}{Jeffrey Pennington},
  \bibinfo{person}{Richard Socher}, {and} \bibinfo{person}{Christopher~D
  Manning}.} \bibinfo{year}{2014}\natexlab{}.
\newblock \showarticletitle{Glove: Global vectors for word representation}. In
  \bibinfo{booktitle}{\emph{Proceedings of the 2014 conference on empirical
  methods in natural language processing (EMNLP)}}.
  \bibinfo{pages}{1532--1543}.
\newblock


\bibitem[Phillips and Ormsby(2016)]%
        {phillips2016industry}
\bibfield{author}{\bibinfo{person}{Ryan~L Phillips} {and} \bibinfo{person}{Rita
  Ormsby}.} \bibinfo{year}{2016}\natexlab{}.
\newblock \showarticletitle{Industry classification schemes: An analysis and
  review}.
\newblock \bibinfo{journal}{\emph{Journal of Business \& Finance
  Librarianship}} \bibinfo{volume}{21}, \bibinfo{number}{1}
  (\bibinfo{year}{2016}), \bibinfo{pages}{1--25}.
\newblock


\bibitem[Platt et~al\mbox{.}(1999)]%
        {platt1999SVM}
\bibfield{author}{\bibinfo{person}{John Platt} {et~al\mbox{.}}}
  \bibinfo{year}{1999}\natexlab{}.
\newblock \showarticletitle{Probabilistic outputs for support vector machines
  and comparisons to regularized likelihood methods}.
\newblock \bibinfo{journal}{\emph{Advances in large margin classifiers}}
  \bibinfo{volume}{10}, \bibinfo{number}{3} (\bibinfo{year}{1999}),
  \bibinfo{pages}{61--74}.
\newblock


\bibitem[Rong(2014)]%
        {rong2014word2vec}
\bibfield{author}{\bibinfo{person}{Xin Rong}.} \bibinfo{year}{2014}\natexlab{}.
\newblock \showarticletitle{word2vec parameter learning explained}.
\newblock \bibinfo{journal}{\emph{arXiv preprint arXiv:1411.2738}}
  (\bibinfo{year}{2014}).
\newblock


\bibitem[Sarmah et~al\mbox{.}(2022)]%
        {sarmah2022learning}
\bibfield{author}{\bibinfo{person}{Bhaskarjit Sarmah}, \bibinfo{person}{Nayana
  Nair}, \bibinfo{person}{Dhagash Mehta}, {and} \bibinfo{person}{Stefano
  Pasquali}.} \bibinfo{year}{2022}\natexlab{}.
\newblock \showarticletitle{Learning Embedded Representation of the Stock
  Correlation Matrix using Graph Machine Learning}.
\newblock \bibinfo{journal}{\emph{arXiv preprint arXiv:2207.07183}}
  (\bibinfo{year}{2022}).
\newblock


\bibitem[Satone et~al\mbox{.}(2021)]%
        {satone2021fund2vec}
\bibfield{author}{\bibinfo{person}{Vipul Satone}, \bibinfo{person}{Dhruv
  Desai}, {and} \bibinfo{person}{Dhagash Mehta}.}
  \bibinfo{year}{2021}\natexlab{}.
\newblock \showarticletitle{Fund2Vec: mutual funds similarity using graph
  learning}. In \bibinfo{booktitle}{\emph{Proceedings of the Second ACM
  International Conference on AI in Finance}}. \bibinfo{pages}{1--8}.
\newblock


\bibitem[Vachhani et~al\mbox{.}(2019)]%
        {vachhani2019machine}
\bibfield{author}{\bibinfo{person}{Hrishikesh Vachhani},
  \bibinfo{person}{Mohammad~S Obiadat}, \bibinfo{person}{Arkesh Thakkar},
  \bibinfo{person}{Vyom Shah}, \bibinfo{person}{Raj Sojitra},
  \bibinfo{person}{Jitendra Bhatia}, {and} \bibinfo{person}{Sudeep Tanwar}.}
  \bibinfo{year}{2019}\natexlab{}.
\newblock \showarticletitle{Machine learning based stock market analysis: A
  short survey}. In \bibinfo{booktitle}{\emph{International Conference on
  Innovative Data Communication Technologies and Application}}. Springer,
  \bibinfo{pages}{12--26}.
\newblock


\bibitem[Wan et~al\mbox{.}(2021)]%
        {wan2021sentiment}
\bibfield{author}{\bibinfo{person}{Xingchen Wan}, \bibinfo{person}{Jie Yang},
  \bibinfo{person}{Slavi Marinov}, \bibinfo{person}{Jan-Peter Calliess},
  \bibinfo{person}{Stefan Zohren}, {and} \bibinfo{person}{Xiaowen Dong}.}
  \bibinfo{year}{2021}\natexlab{}.
\newblock \showarticletitle{Sentiment correlation in financial news networks
  and associated market movements}.
\newblock \bibinfo{journal}{\emph{Scientific reports}} \bibinfo{volume}{11},
  \bibinfo{number}{1} (\bibinfo{year}{2021}), \bibinfo{pages}{1--12}.
\newblock


\bibitem[Weiner(2005)]%
        {weiner2005impact}
\bibfield{author}{\bibinfo{person}{Christian Weiner}.}
  \bibinfo{year}{2005}\natexlab{}.
\newblock \showarticletitle{The impact of industry classification schemes on
  financial research}.
\newblock \bibinfo{journal}{\emph{Available at SSRN 871173}}
  (\bibinfo{year}{2005}).
\newblock


\bibitem[Wu et~al\mbox{.}(2021)]%
        {wu2021equity2vec}
\bibfield{author}{\bibinfo{person}{Qiong Wu}, \bibinfo{person}{Christopher~G
  Brinton}, \bibinfo{person}{Zheng Zhang}, \bibinfo{person}{Andrea
  Pizzoferrato}, \bibinfo{person}{Zhenming Liu}, {and} \bibinfo{person}{Mihai
  Cucuringu}.} \bibinfo{year}{2021}\natexlab{}.
\newblock \showarticletitle{Equity2vec: End-to-end deep learning framework for
  cross-sectional asset pricing}. In \bibinfo{booktitle}{\emph{Proceedings of
  the Second ACM International Conference on AI in Finance}}.
  \bibinfo{pages}{1--9}.
\newblock


\bibitem[Yang et~al\mbox{.}(2018)]%
        {yang2018explainable}
\bibfield{author}{\bibinfo{person}{Linyi Yang}, \bibinfo{person}{Zheng Zhang},
  \bibinfo{person}{Su Xiong}, \bibinfo{person}{Lirui Wei},
  \bibinfo{person}{James Ng}, \bibinfo{person}{Lina Xu}, {and}
  \bibinfo{person}{Ruihai Dong}.} \bibinfo{year}{2018}\natexlab{}.
\newblock \showarticletitle{Explainable text-driven neural network for stock
  prediction}. In \bibinfo{booktitle}{\emph{2018 5th IEEE International
  Conference on Cloud Computing and Intelligence Systems (CCIS)}}. IEEE,
  \bibinfo{pages}{441--445}.
\newblock


\end{thebibliography}

\end{document}